\documentclass[epj]{svjour}
\usepackage{graphics}
\begin{document}
%
\title{Measurements of unpolarized azimuthal asymmetries at COMPASS}
\author{Giulio Sbrizzai \inst{1} on behalf of the COMPASS Collaboration 
}                     
%
%
\institute{Universit\'a degli Studi di Trieste, Dipartimento di Fisica e INFN, Trieste, Italia.} 
%
\date{Received: date / Revised version: date}
%
\abstract{
Azimuthal asymmetries in unpolarized SIDIS can be used to probe the transverse momentum of the quarks inside the nucleon.
 Furthermore they give access to the so-far unmeasured Boer-Mulders function. We report on the extraction of these asymmetries 
 from the COMPASS data taken with a $160$ GeV/c $\mu ^+$ beam impinging on a deuteron target. This asymmetries have been extracted separately for positive and negative hadrons, and the results have been compared with theoretical predictions. 
%
} 
\maketitle
\section{Introduction}
\label{intro}
Semi Inclusive Deep Inelastic Scattering (SIDIS) reactions


are an important tool to probe the structure of the nucleon.

The measurement of azimuthal asymmetries in the unpolarized cross section can give informations about the dynamic of the quarks inside the nucleon.


\begin{figure}
\resizebox{0.50\textwidth}{!}{%
 
 \includegraphics{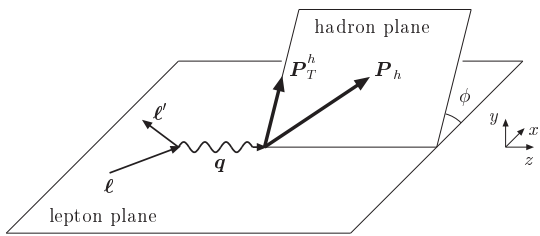}
}
\caption{Definition of the hadron azimuthal angle $\phi$. xz id the lepton scattering plane. $P^h_T$ is the transverse momentum of the inclusively produced hadron with respect to the photon direction.}
\label{fig:1}       
\end{figure}
%

The most general cross section~\cite{Bacchetta}

\begin{eqnarray} 
\frac{\textrm{d}\sigma}{\textrm{d}x\,\textrm{d}y\, \textrm{d}z\, \textrm{d}\phi_h\,P_T^h\,\textrm{d}P_T^h}\, = \\ 
= \frac{\alpha^2}{x y Q^2} \frac{(1+(1-y)^2)}{2} \cdot \nonumber \\               
\cdot   \Big( 1 + \epsilon _1 A_{\cos\phi_h} \cos \phi_h  + \nonumber \\  
+ \epsilon _2 A_{\cos 2\phi_h} \cos2\phi_h + \nonumber \\                 
+ \lambda_l \epsilon _3 A_{\sin\phi_h} \sin\phi_h   \Big)  \;,  \nonumber
\label{cross_section_1}
\end{eqnarray}

has three independent azimuthal modulations: $cos\phi$, $cos2\phi$ and $sin\phi$,
where $\phi$ is the azimuthal angle of the hadron, defined in Fig.~\ref{fig:1}, with respect to the scattering plane. 
Their amplitudes depend on:
\begin{equation}
Q^2 \equiv - q^2\;\;\;
x = \frac{Q^2}{2 \, P \cdot q}, \;\;\;
y =  \frac{P \cdot q}{P \cdot \ell} ,
\;\;\;
z = \frac{P \cdot P_h}{P \cdot q}\,,
\end{equation}
where $\ell$ and $\ell'$ are the 4-momenta of the incoming and outgoing lepton, $P$ is the 4-momentum of the nucleon, $P_h$ the 4-momentum of the outgoing hadron and finally $q = \ell - \ell'$.

The so called kinematical factors are:
\begin{eqnarray}
\epsilon _1 = \frac{2\cdot (2-y)\sqrt{1-y} } {1+(1-y)^2 }\\ \nonumber
\epsilon _2 = \frac{2\cdot (1-y)} {1+(1-y)^2 } \\ \nonumber
\epsilon _3 = \frac {2\cdot y(1-y)}{1+(1-y)^2} \cdot \lambda . \\ \nonumber
\end{eqnarray}

The $sin\phi$ modulation has no clear interpretation in term of Parton Model. 
Its amplitude is proportional to the beam polarization $\lambda$ and it has been found to be small by the CLAS 
collaboration ~\cite{CLAS} which measured an amplitude up to $5\%$ (with a proton target).

The two other amplitudes have been measured in umpolarized SIDIS on proton target 
by previous experiments and results have been published by  EMC ~\cite{EMC_1}, ~\cite{EMC_2}, E665 ~\cite{E665} (in a kinematical region similar to COMPASS) and ZEUS ~\cite{ZEUS}, at higher energies. 
The larger effect has been found for the $cos\phi$ modulation as the European Muon Collaboration measured asymmetries up to 20\%.

The origin of these two modulation is related with the so called
Cahn effect, with the Boer-Mulders Transverse Momentum Dependent Parton Distribution Function (TMD PDF) and with the perturbative QCD corrections to the Quark Parton Model.

The pQCD corrections were first evaluated in ~\cite{Georgi}; later it was shown  
~\cite{Konig} that they are expected to be important only for $P_T^h > 1$ GeV$/c$.
Since the COMPASS kinematical range is such that the most of the statistics is at low $P_T^h$, this contributions 
should be small.

The contribution to the $cos\phi$ and the $cos2\phi$ azimuthal modulation 
given by the Cahn effect comes from kinematics. It is calculated evaluating the elastic quark-lepton 
cross section ~\cite{Cahn}, taking into account the intrinsic transverse momentum of the quark inside the nucleon ($k_T$).
The $cos\phi$ amplitude is proportional to $k_T/Q$ and the $cos2\phi$ amplitude is proportional to $(k_T/Q)^2$.  
In the resulting unpolarized SIDIS cross section, those terms have to be convoluted with the transverse momentum dependent parton distribution functions (TMD PDF) and the fragmentation functions (FF). The first describing the probability to find a quark $q$ with a given $k_T$ and $x$ in the nucleon, the latter describing the probability that the quark $q$ fragments into a hadron $h$ with a given $z$ and $P_T^h$, which is the transverse momentum of the hadron with respect to the direction of $q$, calculated in the GNS.
As predicted in ~\cite{Cahn} the fragmentation proccess should lead to a dilution of the $cos\phi$ and the $cos2\phi$ modulations in the unpolarized SIDIS cross section.

Measurement of the $cos\phi$ term from previous experiment on proton target allowed the extraction of the value of $\langle k_T^2\rangle$ ~\cite{Anselmino_1} 
from the formula:
\begin{eqnarray}
\nonumber &&
\frac{d^5\sigma^{\ell p \to \ell h X }}{dx \, dQ^2 \, dz \, d^2 P_T^h}
\simeq
\sum_q \frac{2\pi\alpha^2e_q^2}{Q^4} \> f_q(x) \> D_q^h(z_h) \biggl[
1+(1-y)^2 \\
&& 
\frac{(2-y)\sqrt{1-y}\> \langle k_T^2\rangle \, z_h \, P_T^h}
{\langle p_t^2\rangle \, Q}\> \cos \phi \biggr]
\frac{1}{\pi\langle P_t ^2\rangle} \, e^{-P_T^h {}^2/\langle p_t^2\rangle} \, ,
\label{cahn-anal-app}
\end{eqnarray}
obtained, considering only the Cahn effect and assuming a gaussian shape for the transverse momentum distributions and neglecting ${\cal O} \left( \frac{k_T^2}{Q^2} \right)$ terms.  
As a result the value for $\langle k_T^2\rangle \simeq 0.25$ (GeV/c)$^2$ was obtained.

Recently ~\cite{BM} it has been shown that the Boer-Mulders (B-M) TMD PDF could give the $cos\phi$ and the $cos2\phi$ azimuthal modulation in unpolarized SIDIS. This so far unmeasured PDF describes the correlation between transverse momentum and transverse spin of a quark inside an unpolarized nucleon and appears in the SIDIS cross section convoluted with the Collins FF. 
In some model calculation ~\cite{Barone} it is predicted that the contribution of the B-M function to the $cos2\phi$ asymmetry is comparable with the Cahn effect but that it significantly differs for positive or negative hadrons.

\section{The COMPASS experiment}
\label{sec:COMPASS}
\begin{figure}
\resizebox{0.50\textwidth}{!}{
\includegraphics{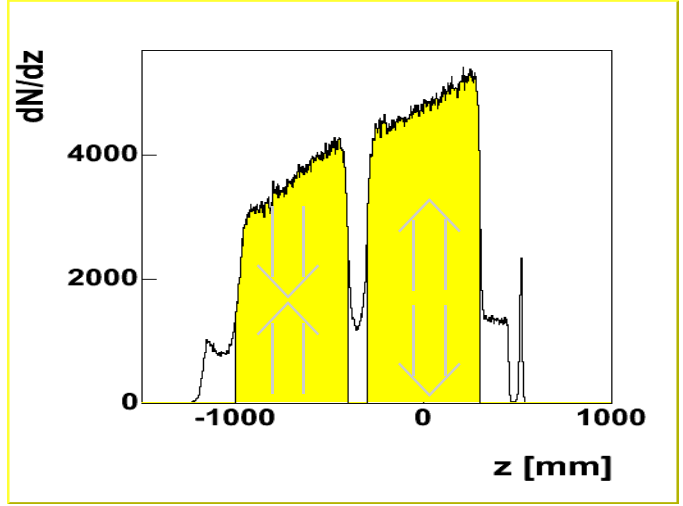}
}
\caption{Interaction vertex distribution. In yellow is highlighted the region corresponding to the $^6LiD$ target cells. The slope, raising going to greater Z values, is due to the target solenoid geometrical acceptance. For each period of data taking, both cells are oppositely polarized.}
\label{fig:target}    
\end{figure}

The COMPASS experiment ~\cite{COMPASS} is a fixed target experiment at SPS at CERN. It features a $160$ GeV$/c \, \mu^+$ beam, with a natural polarization of $-80\%$, and a solid state polarized target. From 2002-2006, data has been taken with a polarized $^6LiD$ target, while in 2007 a $NH_3$ target was used.
The target was either transversely or longitudinally polarized to accomplish the wide experimental program on the spin structure of the nucleons. 
Up to 2004 the target consisted in two separate cells of the same material which were oppositely polarized as shown in Fig. ~\ref{fig:target}, for the transverse polarization. The polarization of the cells have been reversed every week, during the data taking periods, in order to minimize systematic effects.

\subsection{Asymmetries Extraction}
\label{sec:a_extr}
The data sample used for this analysis was taken in the year 2004 with both transversely (T) and longitudinally (L) polarized target. 

The main cuts used in this analysis are:
$Q^2 > 1$ GeV$/c$
$0.1 ~< ~y ~< ~0.9 \,\,\, W ~> ~5 GeV/c^2$, \newline (where $W$ is the mass of the final hadronic state) and:
\newline $0.2 ~< ~z ~< ~0.85$ \,\,\, $0.1 ~< ~p_T^h ~< ~1.5$ ~GeV$/c$. 

The number of positive hadrons after all cuts are:
$3 \cdot 10^6$ and  $2 \cdot 10^6$ respectively in data collected in transverse and longitudinal target spin configuration; 
$2 \cdot 10^6$ and $2 \cdot 10^6$ are the numbers of negative hadrons.

For each target configuration (T,L) data from periods with opposite target polarization have been used to cancel possible terms in the cross section due to the target nucleon polarization.
The data coming from different periods have been weighted accordingly to:
\begin{equation}
\label{eq:weight}
N \, = \, F_2 \cdot P_2 \cdot N_1 \,+\, F_1 \cdot P_1 \cdot N_2,
\end{equation}
where $P_i$ is the mean polarization value, $N_i$ the number of final hadrons, $F_i$ is the measured muons flux; the indexes $1,2$ refer to the data taking periods.

Dedicated Monte Carlo simulations have been performed to correct the hadron azimuthal distributions for the apparatus acceptance, which gives by itself an azimuthal modulation.
Each simulation consisted in the full COMPASS Monte Carlo chain: the SIDIS event generation was performed with LEPTO, the interaction between particles and materials and the detectors response was performed with COMGEANT, a software based on GEANT 3. To perform the tracks reconstructions, the same software (CORAL) was used as for real data. 

The apparatus azimuthal acceptance has been defined as:
$A(\phi) \,=\, N_{rec}(\phi) / N_{gen}(\phi)$,
where $N_{rec}(\phi)$ is the reconstructed hadrons distribution (after all analysis cuts) and $N_{gen}(\phi)$ is the generated hadrons distribution (in the DIS kinematical region). Both were obtained from MC simulations.

The hadron azimuthal distribution corrected by the apparatus acceptance 

$N_{corr}(\phi)\,=\, N(\phi) / A(\phi)$ 

has been fitted with a four parameter function:

$N_{corr}(\phi)\,=\, N_0 \cdot (1\,+\,A^{D}_{cos(\phi)}\cdot cos(\phi)\,+\,A^{D}_{cos(2 \phi)}\cdot cos(2 \phi)\,+\,A^{D}_{sin(\phi)}\cdot sin(\phi))$

where $A^{D}_i$ are the extracted amplitudes of the three independent azimuthal modulation ($A^{D}_{cos\phi} = \epsilon_1 A_{cos\phi}$). 

\begin{figure}
\resizebox{0.40\textwidth}{!}{%
\includegraphics{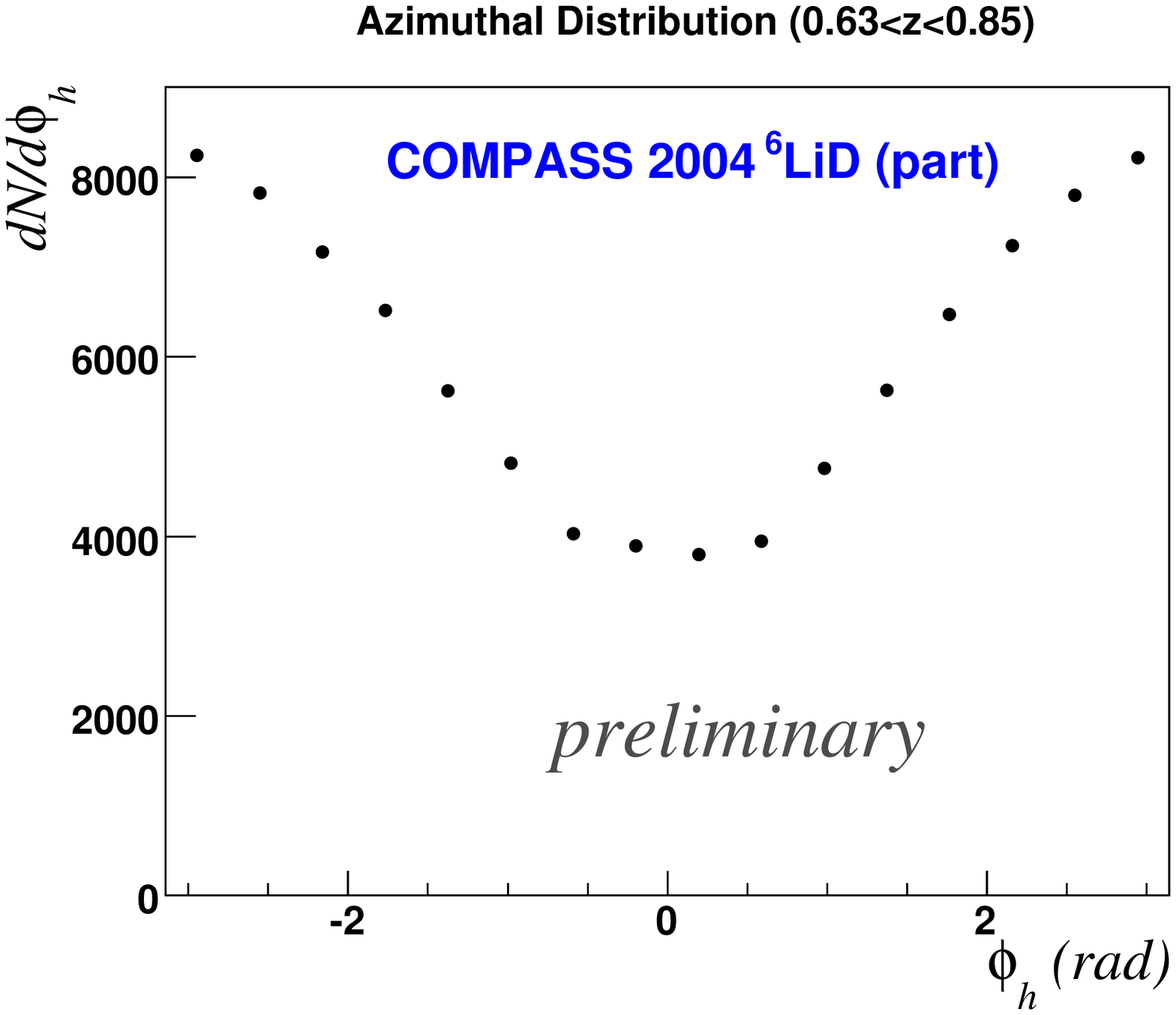}
}
\resizebox{0.40\textwidth}{!}{%
\includegraphics{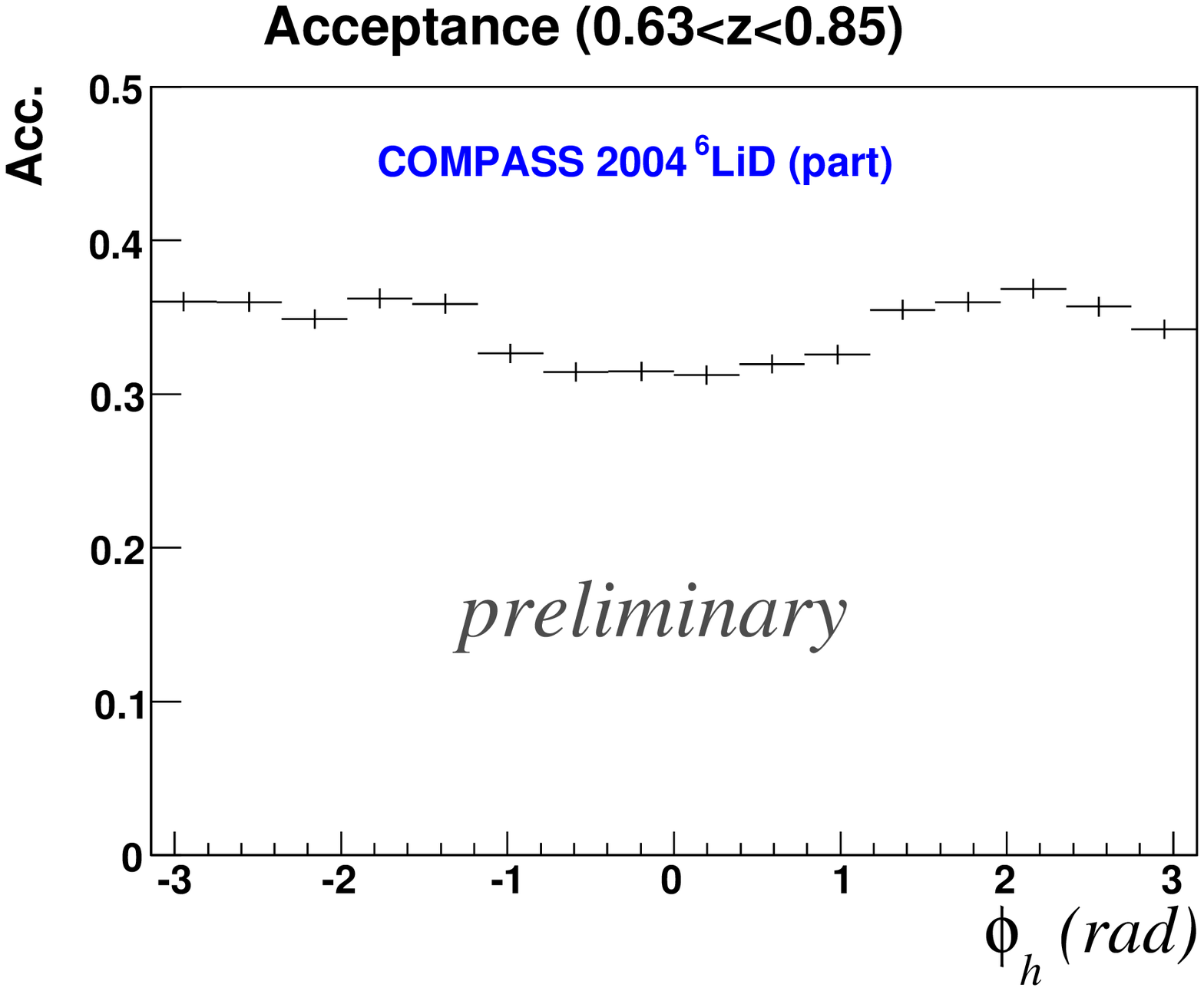}
}
\resizebox{0.40\textwidth}{!}{%
\includegraphics{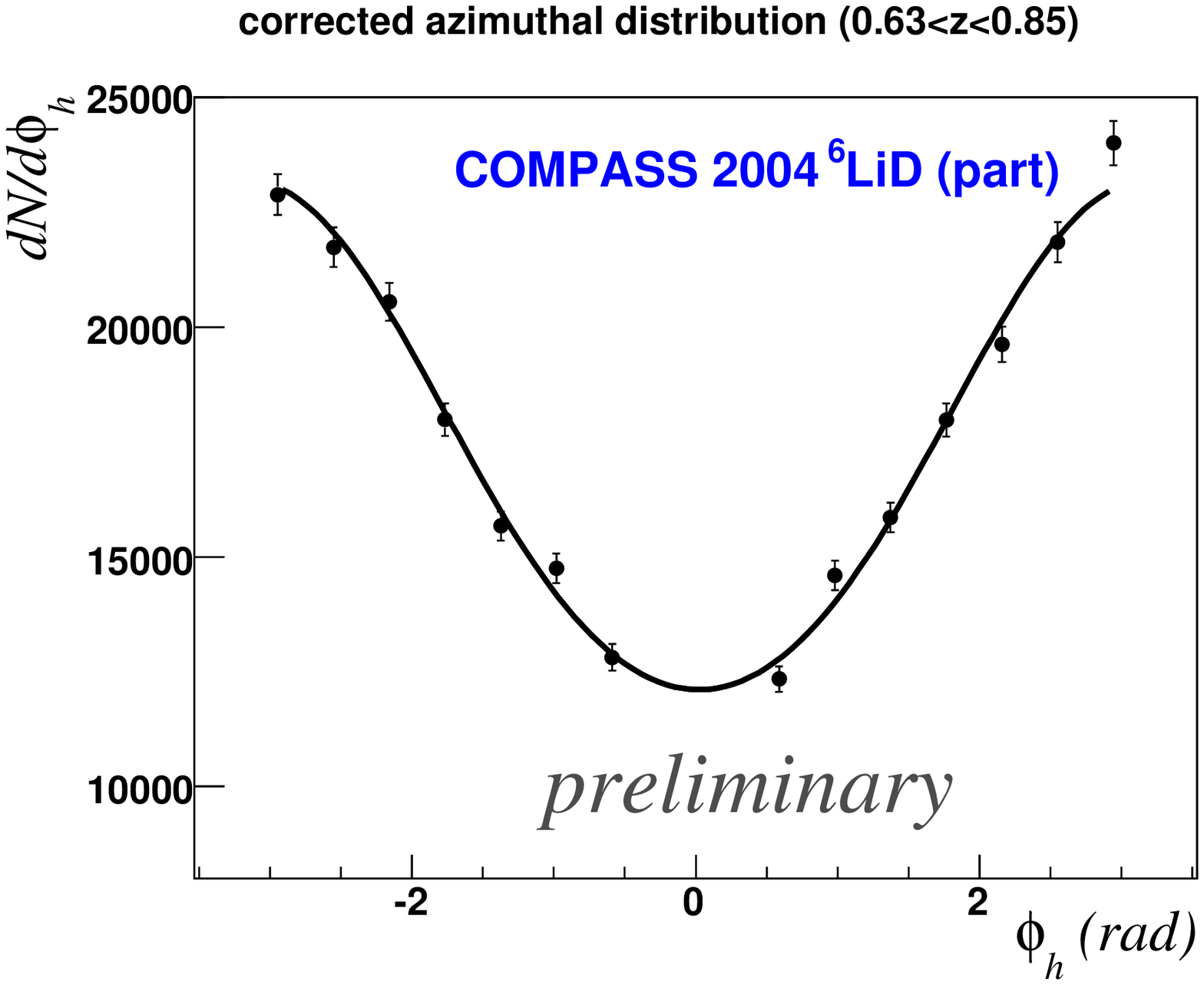}
}
\caption{Top: $N(\phi)$, measured unpolarized azimuthal distribution of hadrons from real data, after weighting over different polarization periods. Middle: $A(\phi)$, acceptance evaluated from the MC simulation. Bottom: $N_{corr}(\phi)$, the corrected hadron azimuthal distribution. All this three distributions correspond to the bin $0.63 < z < 0.85$.}

\label{fig:azi}       
\end{figure}

The asymmetries extraction procedure has been repeated in different bins of $x$, $z$ and $P_T^h$ (averaging over the other two variables), for positive and negative hadrons separately. In Fig.~\ref{fig:azi} tipical hadron azimuthal distributions from real data ($N(\phi)$), the corresponding acceptance from Monte Carlo simulation ($A(\phi)$) and the corrected distribution ($N_{corr}(\phi)$) are shown.

\subsection{Systematic studies}
\label{sec:sys}

Studies were performed to evaluate the contribution of the acceptance corrections to the systematic. The asymmetries were extracted from data taken with longitudinal and transverse target configurations. This means different data taking conditions and apparatus descriptions in the Monte Carlo simulations, mainly concerning the magnetic field in the target zone until the first bending magnet and the beam direction before the target. The asymmetries were extracted using two different sets of LEPTO parameters which described approximately equally well the real data (both for transverse and longitudinal configuration).
All the resulting asymmetries were compared in order to quantify the systematic error in each kinematical bin. 
Further systematic tests performed, like splitting the data sample according to the event topology and time of the measurement, gave no significant contribution.

\subsection{Results and conclusions}
\label{sec:results}

\begin{figure}
\resizebox{0.90\textwidth}{!}{%
\includegraphics{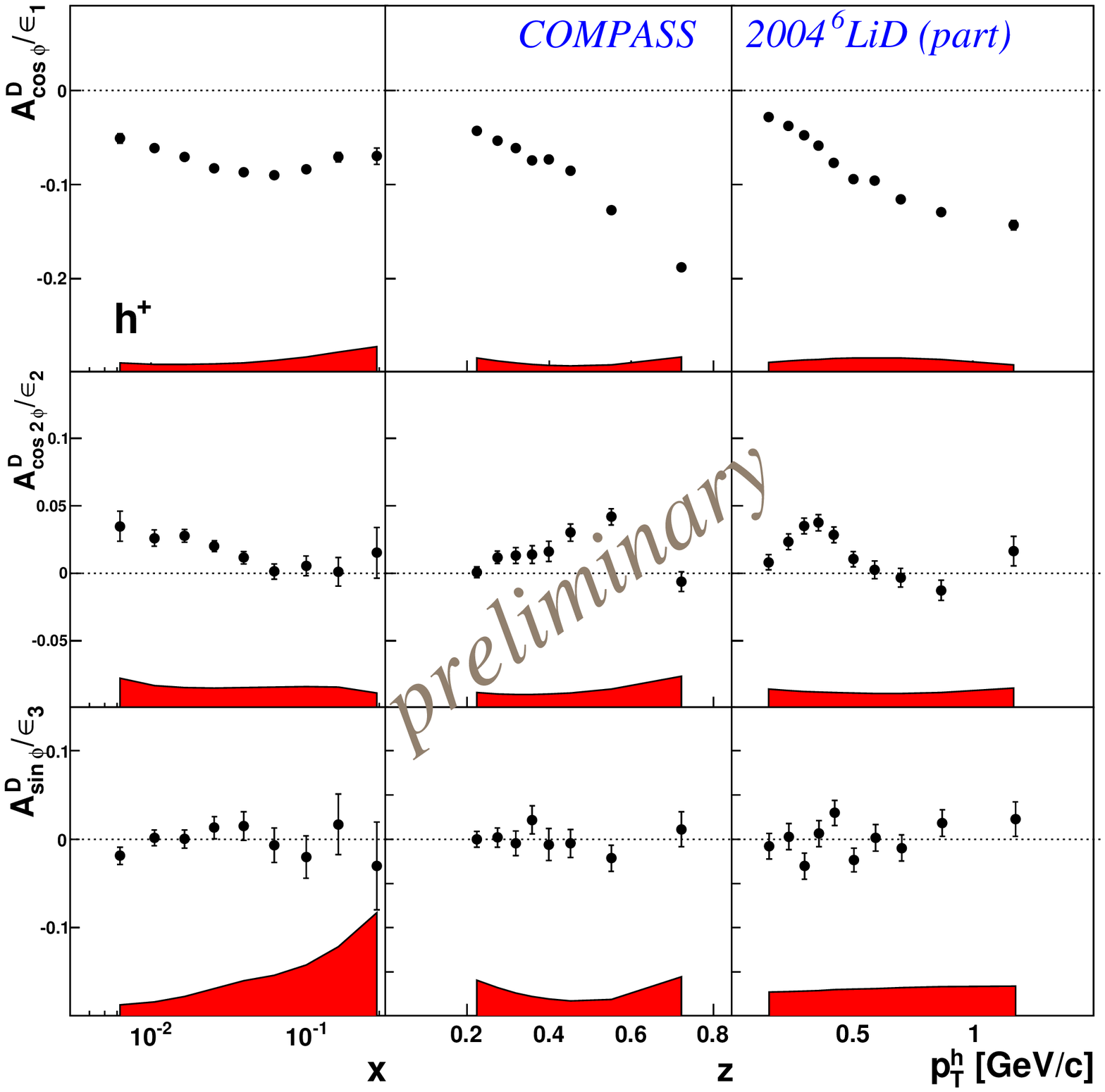}
}
\caption{Asymmetries extracted from positive hadrons as a function of $x$, $z$ and $P_T^h$. The quantities shown are the amplitudes of each modulation divided by the corresponding kinematical factor ($cos\phi$ in the upper raw, $cos2\phi$ in the middle and $sin\phi$, at the bottom)}

\label{fig:pos}       
\end{figure}

\begin{figure}
\resizebox{0.90\textwidth}{!}{%
\includegraphics{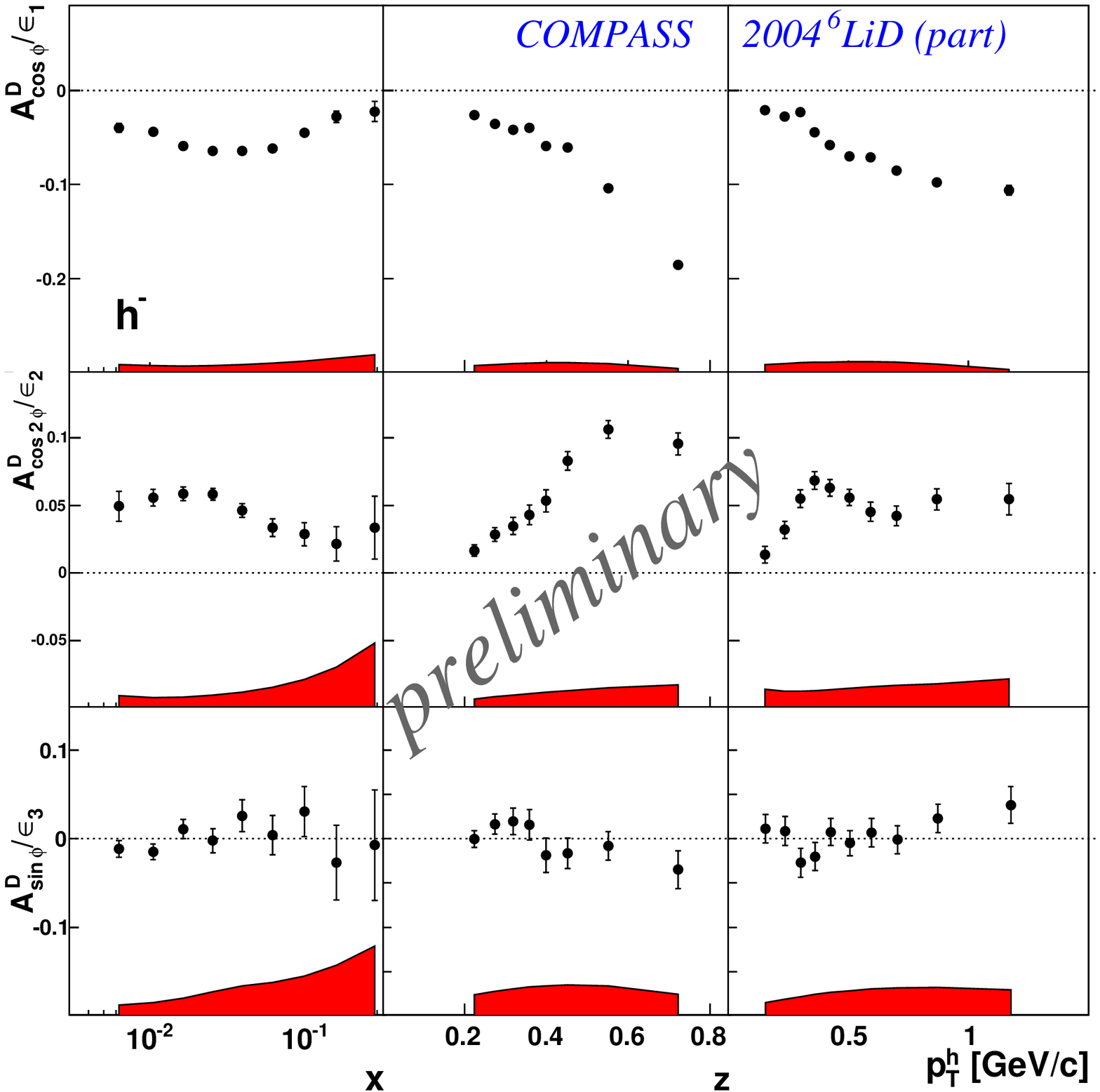}
}
\caption{Asymmetries extracted from negative hadrons as a function of $x$, $z$ and $P_T^h$. The quantities shown are the amplitudes of each modulation divided by the corresponding kinematical factor ($cos\phi$ in the upper raw, $cos2\phi$ in the middle and $sin\phi$, at the bottom)}

\label{fig:neg}       
\end{figure}

The asymmetries extracted from COMPASS deuteron data are shown in Fig. ~\ref{fig:pos} for positive hadrons and in Fig. ~\ref{fig:neg} for negative hadrons.
They have been divided by the corresponding kinematical factors (see section ~\ref{intro}). The (red) bands give the systematic error while the bars give the statistical error (which is small compared with previous experiment).
The $sin\phi$ asymmetries shown in the last line of Fig. ~\ref{fig:pos} and Fig. ~\ref{fig:neg} suggest a beam polarization effect compatible with zero both for positive and for negative hadrons. 
The results for all $cos\phi$ asymmetries (first line) show the largest amplitude and are mostly due to the Cahn effect.

\begin{figure}
\resizebox{0.50\textwidth}{!}{%
\includegraphics{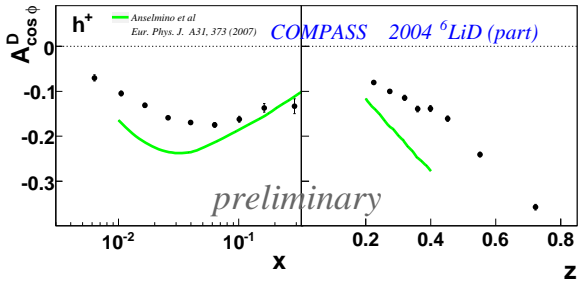}
}
\caption{Comparison with theoretical predictions. Black points are the $cos\phi$ amplitude extracted from COMPASS data, positive hadrons, as a function of $x$ and $z$ and in green are shown the values predicted, in Ref. ~\cite{Anselmino_1}, for COMPASS kinematics without considering the B-M effect.}

\label{fig:th1}       
\end{figure}

A comparison with the theoretical predictions for COMPASS kinematics~\cite{Anselmino_1} is shown in Fig.~\ref{fig:th1}. In~\cite{Anselmino_1} only Cahn effect was considered, being pQCD contribution negligible and the B-M $cos\phi$ mechanism not completely understood. The Cahn effect should not depend on the charge of the hadrons involved, and the differences measured from COMPASS data could hint to a possible B-M PDF contribution.

\begin{figure}
\resizebox{0.70\textwidth}{!}{%
\includegraphics{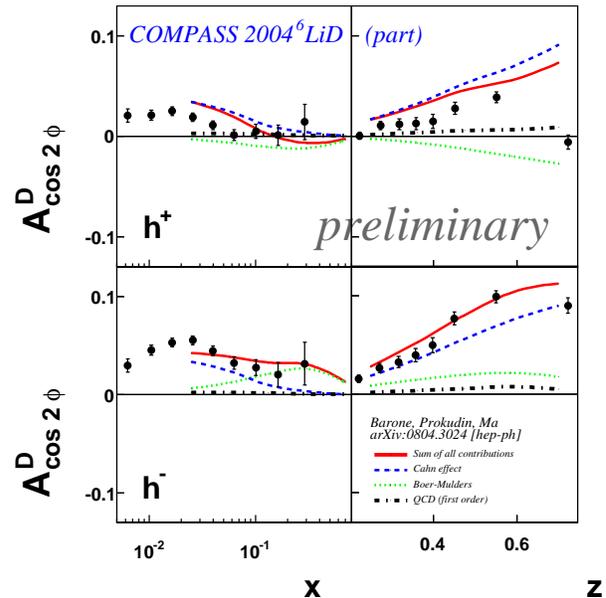}
}
\caption{Comparison with theoretical predictions. Black points are the $cos2\phi$ amplitude extracted from COMPASS data, for positive hadrons (upper two plots) and negative hadrons (lower raw), as a function of $x$ and $z$. The red line shows the predictions from Ref. ~\cite{Barone}, and which is the sum of Cahn contribution (blue line), B-M contribution (green dashed line) and perturbative QCD (black dotted line). These values have been calculated for the COMPASS kinematical region.}

\label{fig:th2}       
\end{figure}

The most evident charge dependence is in the $cos2 \phi$ asymmetry. The results obtained in this analysis are compared to theoretical predictions~\cite{Barone}, made for the COMPASS kinematical region, in Fig.~\ref{fig:th2}. Here different contributions were taken into account: the Cahn effect (blue dashed line in the picture), the B-M PDF (green dashed line) and the first order pQCD contribution which is negligible as shown by the black dashed-dotted line. In this work the B-M PDF was assumed to be proportional to the better known Sivers function. Comparing the results obtained for positive and negative hadrons one can see that only the contribution coming from the B-M PDF changes significantly with the hadron charge. COMPASS measurements confirm this trend and agree quite well with these predictions.

\newpage

\newpage
\clearpage

\clearpage

%
%
%
%

\begin{thebibliography}{}
%
%
\bibitem{Bacchetta}
A.Bacchetta et al.
JHEP \textbf{0702}, (2007) 093
and references therein.

\bibitem{CLAS}
CLAS Collaboration, H.~Avakian et~al. 
Phys. Rev. {\bf D69} (2004) 112004.

\bibitem{Georgi}
R.K.Ellis et al.
Nucl. Phys. \textbf{B152} (1979) 285.


\bibitem{Konig}
  A.~Konig and P.~Kroll,
  Z.\ Phys.\  C {\bf 16}, (1982) 89.

\bibitem{Cahn}
R.N.Cahn, 
Phys. Lett. B \textbf{78} 269, (1978).


\bibitem{Anselmino_1}
M.Anselmino et al.
Phys. Rev. D \textbf{71}, (2005) 074006.


\bibitem{BM}
D.Boer, P.J.Mulders
Phys. Rev. D \textbf{57} (1998) 5780-5786.

\bibitem{Barone}
V.Barone, A.Prokudin, B.Q.Ma
Phys. Rev. D \textbf{78} (2008) 045022. 

\bibitem{EMC_1}
European Muon Collaboration, J.J.Aubert et al. 
Phys. Lett. B \textbf{130}, (1983) 118.

\bibitem{EMC_2}
European Muon Collaboration, M.Arneodo et al.
Z. Phys. C \textbf{34}, (1987) 277.

\bibitem{E665}
E665 Collaboration, M.R.Adams et al.
Phys. Rev. D \textbf{48} (1993) 5057-5066.


\bibitem{ZEUS}
ZEUS Collaboration, J.Breitweg et al.
Phys. Lett. B \textbf{481} (2000) 199-212.

\bibitem{COMPASS}
COMPASS Collaboration, P.Abbon et al.
Nucl. Instr. and Meth. A \textbf{577} (2007) 455-518

\end{thebibliography}
%

\end{document}